\documentclass[format=sigconf,anonymous=false,review=false,screen=true]{acmart}

\AtBeginDocument{%
  \providecommand\BibTeX{{%
    \normalfont B\kern-0.5em{\scshape i\kern-0.25em b}\kern-0.8em\TeX}}}

\copyrightyear{2023} 
\acmYear{2023} 
\setcopyright{acmcopyright}
\acmConference[WWW '23]{Proceedings of the ACM Web Conference 2023}{April 30--May 4, 2023}{Austin, Texas, USA}
\acmBooktitle{Proceedings of the ACM Web Conference 2023 (WWW '23), April 30--May 4, 2023, Austin, Texas, USA}
\acmPrice{15.00}
\acmDOI{10.1145/nnnnnnn.nnnnnnn}
\acmISBN{978-1-nnnn-nnnn-n/nn/nn}

\usepackage{graphicx}
\usepackage[caption=false]{subfig}
\usepackage{color}
\usepackage{algorithm}
\usepackage{algorithmic}

\usepackage{microtype}
\usepackage{amsmath,amsfonts}
\usepackage{mathtools}
\usepackage{mathrsfs}
\usepackage{bm}
\usepackage{booktabs, multirow}
\usepackage{makecell}
\usepackage{stfloats}
\usepackage{xcolor}
\usepackage{enumitem}
\usepackage{enumerate}

\begin{document}

\title{Correlative Preference Transfer with Hierarchical Hypergraph Network for Multi-Domain Recommendation}

\author{Zixuan Xu$^{\ddagger}$, Penghui Wei$^{\ddagger}$, Shaoguo Liu$^{\scriptscriptstyle *}$, Weimin Zhang, Liang Wang and Bo Zheng}
\affiliation{
  \institution{Alibaba Group}
  \country{Beijing, China}
}
\email{{xuzixuan.xzx,wph242967,shaoguo.lsg,dutan.zwm,liangbo.wl,bozheng}@alibaba-inc.com}

\thanks{$^{\dagger}$Co-first authorship. $^{\scriptscriptstyle *}$Correspondence to: S. Liu.}

\renewcommand{\authors}{Zixuan Xu, Penghui Wei, Shaoguo Liu, Liang Wang and Bo Zheng}
\renewcommand{\shortauthors}{Xu and Wei et al.}

\begin{abstract}
Advanced recommender systems usually involve multiple domains (such as scenarios or categories) for various marketing strategies, and users interact with them to satisfy diverse demands. The goal of multi-domain recommendation (MDR) is to improve the recommendation performance of all domains simultaneously. Conventional graph neural network based methods usually deal with each domain separately, or train a shared model to serve all domains. The former fails to leverage users' cross-domain behaviors, making the behavior sparseness issue a great obstacle. The latter learns shared user representation with respect to all domains, which neglects users' domain-specific preferences. 
In this paper we propose $\mathsf{H^3Trans}$, a \textbf{h}ierarchical \textbf{h}ypergrap\textbf{h} network based correlative preference \textbf{trans}fer framework for MDR, which represents multi-domain user-item interactions into a unified graph to help preference transfer. $\mathsf{H^3Trans}$ incorporates two hyperedge-based modules, namely dynamic item transfer (Hyper-I) and adaptive user aggregation (Hyper-U). Hyper-I extracts correlative information from multi-domain user-item feedbacks for eliminating domain discrepancy of item representations. Hyper-U aggregates users' scattered preferences in multiple domains and further exploits the high-order (not only pair-wise) connections  to improve user representations. Experiments on both public  and production datasets verify the superiority of $\mathsf{H^3Trans}$ for MDR.

\end{abstract}

\ccsdesc[500]{Information systems~Personalization}
\ccsdesc[500]{Information systems~Recommender systems}
\ccsdesc[500]{Computing methodologies~Neural networks}

\keywords{Multi-domain Recommendation, Preference Transfer, Hypergraph Learning, Behavior Sparseness}

\maketitle

\section{Introduction}

\begin{figure}[t]
\vspace{2em}

    \centering
    \includegraphics[width=1\columnwidth]{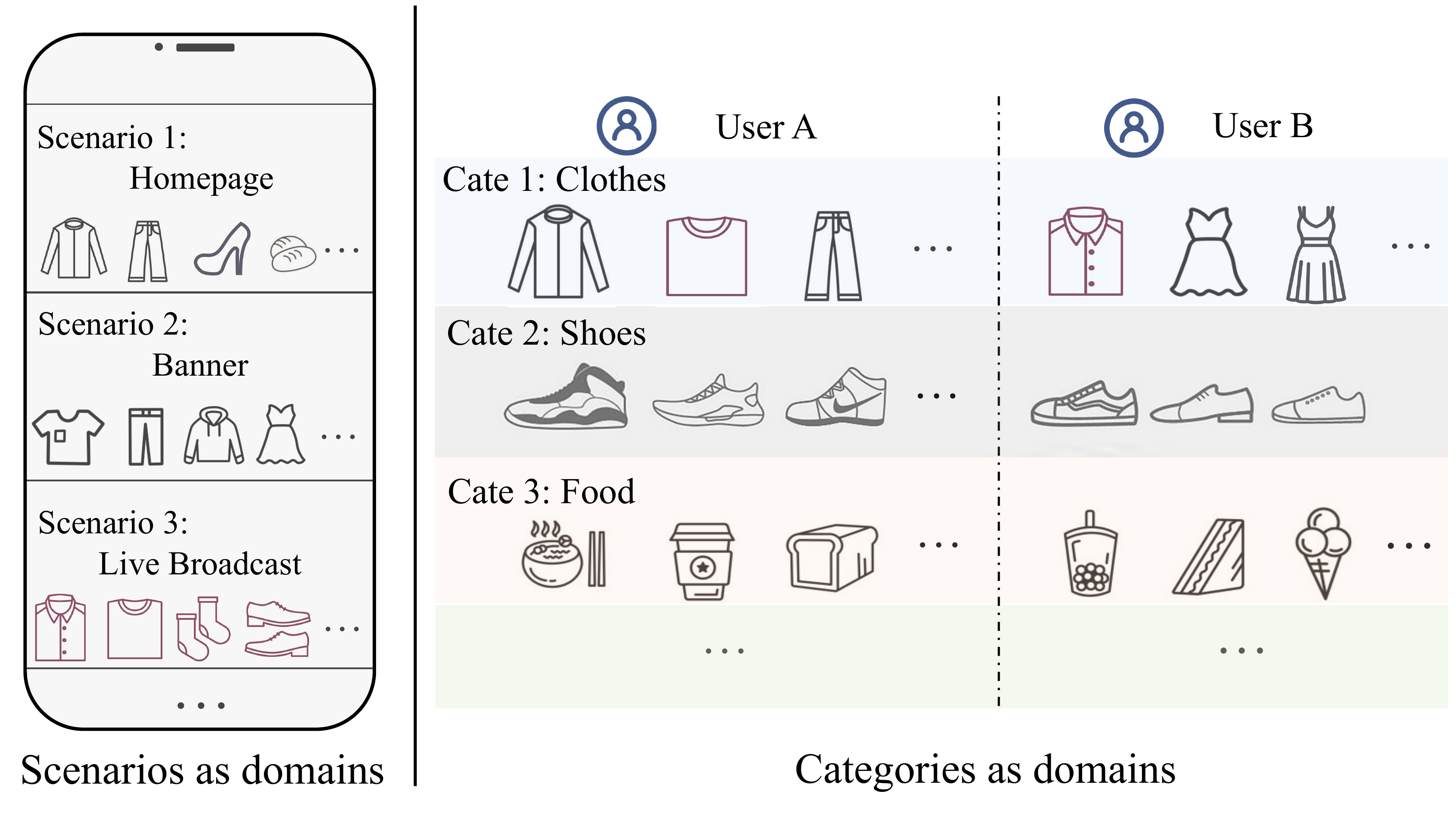}
    \caption{Multi-domain recommendation: the definition of \textit{domain} can be \textit{recommendation scenario} or \textit{item categories}.}
    \label{fig:motivation}
\end{figure}

Personalized recommender systems  aim to make effective and satisfying choices for users. They usually involve multiple recommendation \textbf{scenarios} or domains, and each scenario contains a set of items that is related to the scenario’s topic and marketing strategy. Users interact with these scenarios to satisfy diverse demands. For example, the E-commerce platform \text{Taobao}\footnote{https://www.taobao.com/} provides diversified shopping spots including product search, homepage feed, banner, live broadcast and so on, as shown in the left part of Fig.~\ref{fig:motivation}. \text{Baidu}\footnote{https://www.baidu.com/} serves as a comprehensive website where users can read news, watch videos and more.   Broadly speaking, different item \textbf{categories} can also be regarded as multiple domains. As in the right part of  Fig.~\ref{fig:motivation}, users usually interact with various categories such as  clothes, food and more for their different demands.

\textbf{Multi-domain recommendation} (MDR) has attracted increasing research attention, the goal of which is to improve the recommendation performance of all domains simultaneously. 
There are both commonality and diversity among domains. For the commonality, multiple domains usually have common users and overlapped items, and a user may have similar behavior patterns across domains (for example,  preferring ordinary or fashionable goods). The users' domain-invariant preference and items' static information can be shared across domains. 
For the diversity, the domains have different topics with specific items, thus attract different audiences and cause discrepant data distributions. 

Graph neural networks (GNNs) have proven to be powerful for recommendations because user-item interactions are naturally suitable for modeling as a graph.
Conventional GNN-based methods for MDR can be divided into two types. The first type deals with domains separately. That is, for each domain we construct user-item interaction graph and train model independently, which learns separate representations for different domains to characterize users' domain-specific preferences. However, the sparseness of interaction behaviors in emerging domains~\cite{gu2021self,chen2020scenario} is a crucial obstacle.
The second type alternatively constructs a unified interaction graph using multi-domain data and train a shared model to serve all domains~\cite{PPGN}. 
Considering the intrinsic difference among domains' data distributions, the shared model neglects domain-specific characteristics which results in limited performance.

Researchers have proposed some advanced methods ~\cite{pretrain,BiTGCF,DAN,dagcn,da-gtcdr} that exert the prominent feature extracting ability of GNN and incorporate knowledge transfer to alleviate the sparseness. For example, pretrain-finetune diagram which transfers a pre-trained graph encoder to initialize the node embedding on the target domain is a widely used way~\cite{pretrain}. Considering the pretrain-finetune paradigm only improves the recommendation accuracy on a single target domain, some works exploit to improve the recommendation accuracy on both domains simultaneously~\cite{BiTGCF,da-gtcdr}.
Despite their effectiveness, these methods focus on knowledge transfer between only two domains. When employed in more than two domains, they only capture pair-wise relations between domains and dismiss the high-order connections.

For effective MDR, the key is to learn from the interactions in all domains and acquire transferable knowledge to obtain better user representations that characterize their domain-specific preferences. 
In this paper we propose $\mathsf{H^3Trans}$, a \textbf{h}ierarchical \textbf{h}ypergrap\textbf{h} network based correlative preference \textbf{trans}fer framework to improve MDR. As a general topological structure, a hyperedge can connect an arbitrary number of nodes, and thus hypergraph provides a means for modeling high-order connections in multiple domains. 
We integrate users' multi-domain behaviors into a unified graph and incorporate hyperedges to help preference transfer. Specifically, each user is viewed as multiple nodes w.r.t. to different domains, where the representation of each user node characterizes the domain-specific preference. For item nodes, because items' properties are relatively static than users, we view each item as a single node shared by all domains. 

The core of the hypergraph structure constructed by $\mathsf{H^3Trans}$ is two novel types of hyperedges for improving user and item representation learning. We first design a dynamic item transfer module named Hyper-I. For a given domain, we dynamically seek out related items from user-item interactions of other domains, and construct a hyperedge (named hyperedge-i) to connect them as cross-domain item relations. Hyperedge-i helps build relations between the items of different domains and capture users' correlative preference from the cross-domain behaviors without interference information. Moreover, we propose a structure-aware aggregator with attention mechanism to model the message passing procedure through hyperedge-i, which adjusts item representations much more correlative to the target domain and thus improves the recommendation performance in multiple domains. 

We further introduce an adaptive user aggregation module named Hyper-U. Each user is viewed as a separate node per domain, that is, for a given user we can acquire separate user representations in multiple domains.
We utilize a hyperedge (named hyperedge-u) to connect these separate user nodes of a given user, which aggregates the scattered user preferences among multiple domains. To effectively model the high-order connections among domains, we propose to employ attention mechanism into the message propagation within such hyperedges. Hyperedge-u contributes to transferring correlative preferences from source domains and capturing the commonality among multiple domains. Note that each domain can be viewed as the target domain (and the others as the sources), thus our proposed $\mathsf{H^3Trans}$ can improve the quality of user representation for all domains simultaneously. 

The contributions are as follows:
\begin{itemize}
    \item We propose $\mathsf{H^3Trans}$, a hierarchical hypergraph network based correlative preference transfer framework for MDR. 
    To our knowledge, this is the first work that investigates hypergraph-based preference transfer in MDR.
    \item To improve item representations for cross-domain transfer, Hyper-I performs dynamic item transfer which helps extract correlative preference from the cross-domain behaviors without interference information. 
    \item To model the high-order connections among users' multi-domain behaviors, Hyper-U aggregates users' scattered preferences in multiple domains and exploits the high-order connections with an attention based propagation layer.  
    \item Extensive experiments on large-scale production datasets and public datasets are conducted to analyze our proposed $\mathsf{H^3Trans}$, and the results demonstrate the superiority.
\end{itemize}

\section{Preliminary}
\subsection{Definition of Hypergraph}
Compared to an ordinary graph, a hypergraph is a more general topological structure where a \textbf{hyperedge} can connect an arbitrary number of nodes. Formally, a hypergraph is composed of a node set and a hyperedge set. The connectivity of a hypergraph can be represented by an incidence matrix ${H}$, where $h_{ve}=1$ if the hyperedge $e$ contains the node $v$, otherwise $h_{ve} = 0$. Besides, we use $E_v$ to denote a set of hyperedges that connect to node $v$, and use $V_e$ to denote a set of nodes connected to hyperedge $e$. Also, we can define the neighbors $\mathcal{N}_v$ of node $v$ as a set of nodes that share at least one hyperedge with node $v$.

\subsection{Problem Definition}
Given domains $\left\{\mathcal{D}_m\right\}_{m=1}^T$, where $T$ denotes the number of domains. 
For domain $\mathcal{D}_m$, we utilize $\bm{U}^m$ and $\bm{I}^m$ to denote its user ID set and item ID set respectively. 
Let $\mathcal{R}^m \in \mathbb{R}^{|\bm{U}^m| \times |\bm{I}^m|}$ denotes the user-item interaction matrix of domain $\mathcal{D}_m$. If its entry $r_{ui}^m = 1$, it means that the user $u$ interacted with the item $i$ under domain $m$. In this work, we consider click behavior as the interaction type. 

Given a specific domain $\mathcal D_m$, the problem of single-domain recommendation is to estimate the scores of unobserved entries in one interaction matrix $\mathcal{R}^m$, and we compute the score between a user and an item as:
\begin{equation}
{\hat{r}}^m_{u, i} = f(z_u, z_i \mid \mathcal{D}_m)    
\end{equation}
Here $z_u$ and $z_i$ denote the learned representations of user $u\in\bm U^m$ and item $i\in\bm I^m$ for domain $\mathcal{D}_m$, and $f(\cdot)$ is the similarity function. 

The problem of multi-domain recommendation is to estimate the unobserved scores for all interaction matrices $\{\mathcal{R}^m\}_{m=1}^T$. Specifically, the \textbf{user set} $\bm U$ is shared among all $T$ domains, i.e., $\bm{U} = \bm{U}^1 = \bm{U}^2 = \dots = \bm{U}^T$, because each user may actively interact with all domains. For the \textbf{item set} $\bm I$, each domain has its own set and we denote the total item candidate pool as $\bm{I} = \bm{I}^1 \cup \bm{I}^2 \cup \dots \cup \bm{I}^T$. Note that different domains may have overlapped items.

\begin{figure*}[t]
    \centering
    \centerline{\includegraphics[width=2\columnwidth]{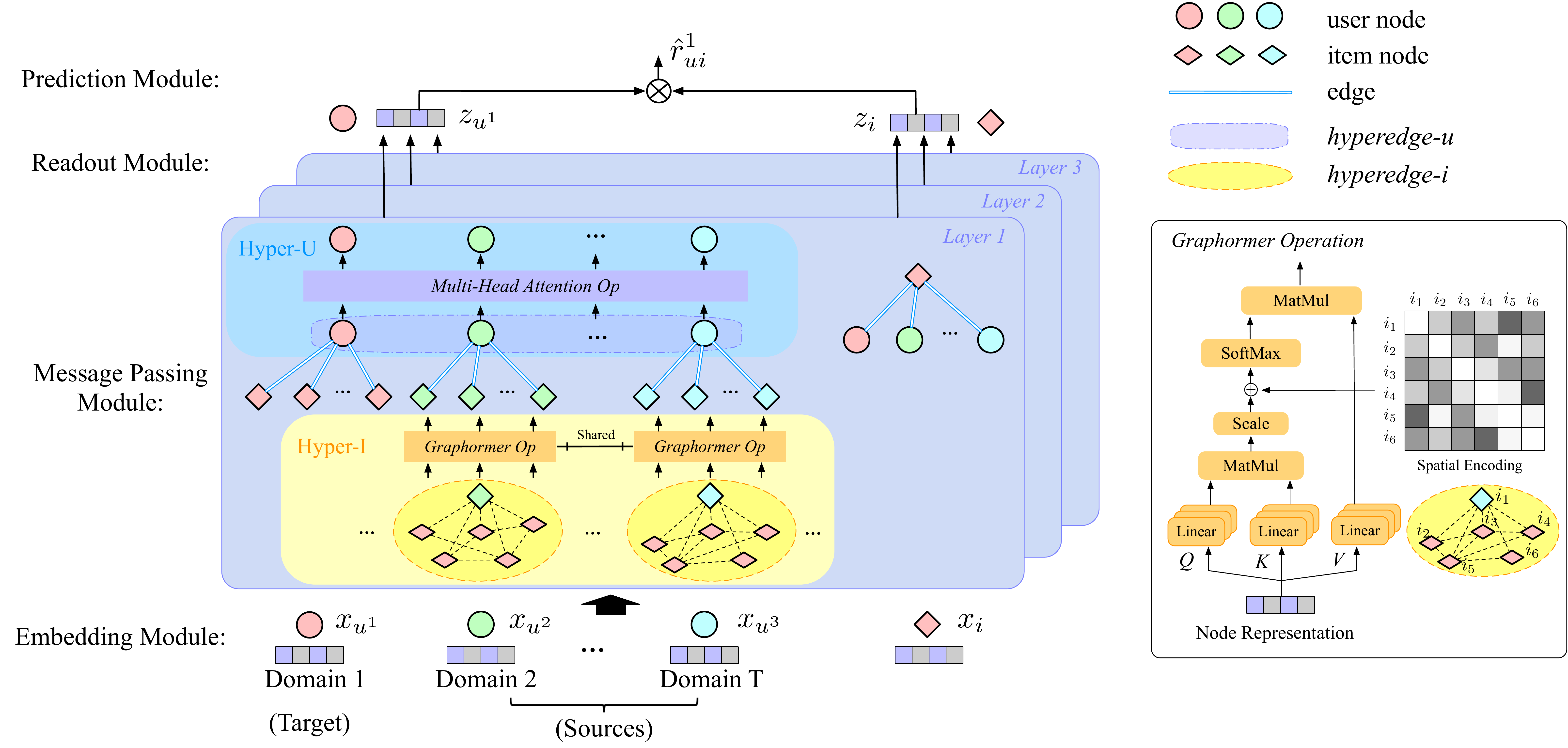}}
    \caption{Overall architecture of $\mathsf{H^3Trans}$. It contains two hyperedge-based modules: adaptive user aggregation (Hyper-U) and dynamic item transfer module (Hyper-I). These two modules compose a hierarchical hypergraph neural network. Different colors refer to different domains. Here we regard the first domain $\mathcal{D}_1$ as target domain and the others are sources.}
    \label{fig:overview}
\end{figure*}

\section{Methodology}
Fig.~\ref{fig:overview} shows the overall architecture of $\mathsf{H^3Trans}$. 
We introduce the construction of multi-domain graph, and basic graph neural network in §~\ref{hypergraph_construction} and \ref{basic_gnn}. Two core modules, namely dynamic item transfer and adaptive user aggregation, compose a hierarchical hypergraph neural network, are introduced in §~\ref{hyper-i} and \ref{hyper-u}. Finally, §~\ref{hypergraph_train} gives training procedure and optimization.

\subsection{Unified Multi-domain Graph}\label{hypergraph_construction}
To improve recommendation performance in all domains, instead of constructing individual graph for each domain, we integrate users' multi-domain behaviors into a unified graph $\mathcal{G}=(\mathcal{V}, \mathcal{E})$.
In details, the node set $\mathcal{V}$ consists of  user nodes and item nodes, i.e., $\mathcal{V} = \mathcal{U} \cup \mathcal{I}$. For \textbf{user nodes}, considering the domain discrepancy and the diversity of users' multi-domain behaviors, it is necessary to acquire separate representations for different domains. Thus we regard each user as separate nodes positioned in different domains. Specifically, for a given user $u\in \bm U$, it corresponds to $T$ nodes $(u^1, u^2, \cdots, u^T)$, thus the relation between user node set size $|\mathcal U|$ and user ID set size $|\bm U|$ meets the condition of $|\mathcal U| = |\bm U|\cdot T$. Each user node representation characterizes user's preference under a specific domain. Then for \textbf{item nodes}, items' properties are relatively static than users. Thus we treat each item $i\in \bm I$ as a single node across various domains. In other words, each item $i$ only corresponds to one node in the graph. The item $i$'s node is also denoted as $i$. 

The \textbf{basic edge set} collects the user-item history interactions from all domain, i.e.\ , $\mathcal{R} = (\mathcal{R}^1, \mathcal{R}^2, \cdots, \mathcal{R}^T)$, where $\mathcal{R}^m$ denotes the user-item interaction matrix of domain $\mathcal{D}_m$. This work considers click behavior as the interaction type. 
For an entry $r_{ui}^m=1$, it means that the user $u$ has interacted with the item $i$ under domain $\mathcal{D}_m$, and we build an interaction edge between the corresponding user node $u^m$ and item node $i$, denoted as $e(u^m, i)$. To clarify which domain the edges belong to, we utilize distinct edge types for different domains. For domain $\mathcal{D}_m$, the edge subset is denoted as $\mathcal{E}^m$, and the whole edge set is the union of all domains as well as hyperedges, i.e.,  $\mathcal{E} = \mathcal{E}^1  \cup \mathcal{E}^2  \cup \cdots  \cup \mathcal{E}^T \cup \mathcal{E}_\mathrm{hyper}$. We detail the construction of \textbf{hyperedge set} in §~\ref{h3}. 
With access to user-item interactions in any domain, it is convenient to leverage hyperedges to build cross-domain relations and capture correlative knowledge.

\subsection{Basic Graph Neural Network}\label{basic_gnn}
We first introduce a base GNN that learns node representations without considering multi-domain relationships. The base GNN includes four modules: (1) embedding module that transforms nodes' sparse attribute features into low-dimensional embeddings; (2) message-passing module with several layers that learn node representations by aggregating information from neighbors; (3) readout module that generates nodes' final representation; (4) prediction module that produces prediction score.

\subsubsection{\textbf{Embedding Module}}
This module maps each node into a $d$-dimensional embedding vector $x_{u^m}$ (or $x_i$). For each user node $u^m\in \mathcal{U}$ (or item node $i \in \mathcal{I}$), we acquire its embedding $x_{u^m}$ (or $x_i$) from a learnable look-up table $X \in \mathbb{R}^{(|\mathcal{U}| + |\mathcal{I}|)\times d}$. Each user corresponds to $T$ nodes, and these nodes share the same initial embedding. Note that each user has attribute features, and the corresponding nodes share the same attribute embedding.

\subsubsection{\textbf{Message Passing Module}}
The message-passing module consists of several layers that follow the neighborhood aggregation scheme. It can be taken as a two-stage process to learn node representations by aggregating information from neighbors. The two stages are neighbor aggregation and node update:

\textbf{Neighbor aggregation}:
\begin{equation}
    \begin{split}
        h_{\mathcal{N}_{u^m}}^{(l)} & = \mathrm{AGG_U}\left(\left\{h_i^{(l-1)} \mid   i \in \mathcal{N}_{u^m} \right\}\right) \\
        h_{\mathcal{N}_i}^{(l)} & = \mathrm{AGG_I}\left(\left\{h_{u^m}^{(l-1)} \mid   {u^m} \in \mathcal{N}_i \right\}\right) \\
    \end{split}
\end{equation}

\textbf{Node update}:
\begin{equation}
    \begin{split}
        h_{u^m}^{(l)} &= \mathrm{UP_U} \left(h_{u^m}^{(l-1)}, h_{\mathcal{N}_{u^m}}^{(l)} \right)  \\
        h_i^{(l)} &= \mathrm{UP_I} \left(h_i^{(l-1)}, h_{\mathcal{N}_{i}}^{(l)} \right) 
    \end{split}
\end{equation}
where $l$ denotes the $l$-th message passing layer. $h_{u^m}^{(l)}$ and $\ h_i^{(l)}$ refer to the hidden representation of user node ${u^m}$ and item node $i$ respectively. $\mathrm{AGG_U}$ and $\mathrm{AGG_I}$ are the aggregation functions for user and item nodes. The same is to the node update function $\mathrm{UP_U}$ and $\mathrm{UP_I}$. There are a lot of designs for aggregate and update function. Here we use mean pooling for the aggregator and linear transforming for node update. Noted that the initial representation is acquired from embedding module, i.e., $h_{u^m}^{(0)}=x_{u^m}, h_i^{(0)}=x_i$.

\subsubsection{\textbf{Readout Module}}
After obtaining $L$ layers representations, we utilize a readout layer to generate the final representation:
\begin{equation}
    z_v  =  \mathrm{Readout} \left(h_v^{(l)} \mid l \in [1, \dots, L]\right),
\end{equation}
where the subscript $v$ can denote user node $u^m$ or item node $i$. Common designs for the readout function include last-layer only, concatenation, and weighted sum. Here we adopt last-layer only. 

\subsubsection{\textbf{Prediction Module}}
The prediction module produces the prediction score that how likely a user $u$ would interact with item $i$ under domain $\mathcal{D}_m$. It is formulated as:
\begin{equation}
    \hat{r}^m_{u, i} = f(z_{u^m}, z_i)
\end{equation}
where $f$ is the score function and we usually adopt similarity function such as inner product and cosine function. 

\subsection{Hierarchical Hypergraph Network}\label{h3}
The base GNN cannot model the multi-domain relation well. We propose $\mathsf{H^3Trans}$ which utilizes hyperedge to exploit high-order connections among users' multi-domain behaviors with two hyperedge-based modules: dynamic item transfer module (Hyper-I) \& adaptive user aggregation module (Hyper-U). These two modules have a hierarchical connection structure and compose a hierarchical hypergraph neural network.

\subsubsection{\textbf{Hyper-I: Dynamic Item Transfer Module}} \label{hyper-i}
In MDR, each domain contains a set of items that is related to the domain's topic and marketing strategy. Due to the intrinsic difference, directly transferring users' cross-domain behaviors from multiple sources to the target is not a good approach. It will introduce interference information and degenerate user representations. To extract correlative preference from users' cross-domain behaviors, we design a dynamic item transfer module, namely Hyper-I. It dynamically adjusts source item representations during transfer to be more relevant to a given target domain, that contributes to capturing correlative user preferences from sources. 

Take domain $\mathcal D_t$ as target domain, and the others as source domains. For each source domain $\mathcal D_s$, before feeding item node hidden representation $h^{(l)}_{i}$ into message passing layers that acquire user node representation by aggregating information from neighboring items, we adjust the item representations to eliminate domain discrepancy. 
Specifically, for each user's interacted item under source domain $\mathcal D_s$, we seek out similar items from the target domain $\mathcal D_t$, and then construct a hyperedge (named \textbf{hyperedge-i}) to connect these item nodes. This hyperedge contains a two-level relationship. The first level is that the interacted source item is related to the picked target items. The second level is that the picked target items are also related to each other. We first introduce the method to seek out the related target items, and then we design a structure-aware hypergraph layer to adjust item representations. 

\paragraph{\textbf{Hyperedge Construction.}}
For a given interacted item $i$ in a source domain $\mathcal D_s$, we seek out a similar item set $\mathcal S_i^t$ from the target domain $\mathcal D_t$, and construct a hyperedge to connect the source item node $i$ and the item nodes of picked item set $\mathcal S_i^t$.
We offer two ways to get similar items: path-based and embedding-based. 

\begin{itemize}
    \item Path-based: Utilize co-occurrence relation among items. We assume that: if there is a user $u$ that clicked on both items $i$ and $j$, then the two items are similar. We design a walk path $(i\rightarrow u^s \rightarrow u^t \rightarrow j)$ and sample $k$ items from the item set $\mathcal I^t$ of target domain as similar items. To avoid noisy paths, we restrict that the click timestamp of each item in $\mathcal S_i^t$ should lie in a range of the source item $i$'s click timestamp. 
    \item Embedding-based: Path-based method is an intuitive way but it seriously relies on the interaction history of users. Embedding-based method makes use of the hidden representation of items $h_i^{(l-1)}$. It leverages the appropriate nearest neighbor algorithm to find the top-$k$ similar items from the target domain, where the source item node $i$ is query and $\mathcal I^t$ is candidate set.
\end{itemize}

\paragraph{\textbf{Graphormer Layer.}}
To perform message passing within the hyperedge, UniGNN~\cite{unignn} and AllSet~\cite{Chien2022YouAA} propose a message passing paradigm on the hypergraph. UniGNN rethinks the message-passing layer of the basic GNN as a two-stage aggregation process. In the first stage, for each hyperedge, use a permutation-invariant function to aggregate the information of the nodes within it. In the second stage, update each node with its incident hyperedges using another aggregating function. The method of AllSet is similar. 

We claim that the above message-passing paradigm fails to model the two-level relationship within {hyperedge-i}. Instead, we employ attention~\cite{vaswani2017attention} to adjust the item representation. Moreover, to effectively exploit the topology structure within the {hyperedge-i}, we introduce the distance matrix of the shortest path among the picked nodes (denoted as $\bm B$) into the attention layers, as introduced in \cite{graphormer}. Fig.~\ref{fig:overview} illustrates the details of this module. Specifically, it refines ${h}_i^{(l-1)}$ before neighbor aggregation:
\begin{equation}
    {h}_i^{(l-1)} \leftarrow \mathrm{GH}_{\mathrm{HyperI}}\left(\text{Concat}\left(h_i^{(l-1)}, \left\{h_j^{(l-1)} \mid | j\in \mathcal S_i^t\right\}\right)\right)[0] 
\end{equation}
where $\mathrm{GH}_{\mathrm{HyperI}}(\cdot)$ is the graphormer layer for Hyper-I module:
\begin{equation}
\small
    \begin{split}
        \mathrm{GH}_{\mathrm{HyperI}} (H_{\mathrm{I}}) & = \text{Concat}\left(\mathrm{Attn}_{\mathrm{I},1}(H_{\mathrm{I}}), \cdots, \mathrm{Attn}_{\mathrm{I},P}(H_{\mathrm{I}}) \right) W_{\mathrm{I}}^O , \\
        \mathrm{Attn}_{\mathrm{I},p}(H_{\mathrm{I}}) &= \mathrm{softmax}\left(\frac{Q_{\mathrm{I},p} {K_{\mathrm{I},p}} ^{\top}}{\sqrt{d_{h_i}/P}} + \Phi({\bm B}) \right)V_{\mathrm{I},p}, \\
        Q_{\mathrm{I},p} & = H_{\mathrm{I}} W_{\mathrm{I},p}^Q, K_{\mathrm{I},p} = H_{\mathrm{I}} W_{\mathrm{I},p}^K, V_{\mathrm{I},p} = H_{\mathrm{I}} W_{\mathrm{I},p}^V \\
    \end{split}
\end{equation}
here $\Phi()$ is a learnable function shared across all layers that maps the distance between every paired nodes to a scalar. $W_*^Q, W_*^K, W_*^V$, and $W_*^O$ are training parameters.

\subsubsection{\textbf{Hyper-U: Adaptive User Aggregation Module}}\label{hyper-u}
After adjusting item representations with Hyper-I, we acquire the representations of separate user nodes by aggregating adjusted representation of their neighbor items. Each user corresponds with multiple nodes that characterize the user's domain-specific preference. Next step is to transfer correlative user preferences from source domains to the target and refine the user representation of target domain. 

Noted that the preference transfer in MDR involves more than one source. The key point is how to aggregate users' scattered preferences in multiple domains and adequately exploit the high-order connections among them. Here we integrate a hyperedge-based module: Hyper-U, to realize adaptive user aggregation.

\paragraph{\textbf{Hyperedge Construction}}
We utilize hyperedge to connect nodes that belong to the same user, and we name this hyperedge as \textbf{hyperedge-u}. Within {hyperedge-u}, each separate node representation characterizes user's interest preference under a specific domain. The {hyperedge-u} connects these separate user nodes and bridges the information propagation across domains, thus realizing adaptive preference transfer. Moreover, benefiting from that hyperedge connects plural nodes, {hyperedge-u} can further exploit the high-order (more than pairwise) connections among multiple domains.

\paragraph{\textbf{Multi-head Attention Layer.}}
We design a new message passing layer for the {hyperedge-u} to replace the original layer. For the $l$-th layer, we first acquire user's separate representations under multiple domains, denoted as $[h_{u^1}^{(l)}, h_{u^2}^{(l)}, \cdots, h_{u^T}^{(l)}]$. Hyper-U module take these separate representations as input, and then refine these representations by aggregating users' scattered preferences and transferring knowledge from other domains. Considering the domain discrepancy and diversity of users' multi-domain behaviors, we employ self-attention mechanism in the Hyper-U module to adaptively fuse users' cross-domain interest representations. Specifically, it refines representations after node update:
\begin{equation}
    \left [{h}_{u^1}^{(l)}, {h}_{u^2}^{(l)}, \cdots, {h}_{u^T}^{(l)}\right ] \leftarrow \mathrm{MHA}_{\mathrm{HyperU}}\left(\left [h_{u^1}^{(l)}, h_{u^2}^{(l)}, \cdots, h_{u^T}^{(l)}\right ]\right), 
\end{equation}
where $\mathrm{MHA}_{\mathrm{HyperU}}(\cdot)$ denotes the multi-head attention layer:
\begin{equation}
\small
    \begin{split}
        \mathrm{MHA}_{\mathrm{HyperU}}(H_{\mathrm{U}}) & = \text{Concat}\left(\mathrm{Attn}_{\mathrm{U},1}(H_{\mathrm{U}}), \cdots, \mathrm{Attn}_{\mathrm{I},P}(H_{\mathrm{U}}) \right) W_{\mathrm{U}}^O , \\
        \mathrm{Attn}_{\mathrm{U},p}(H_{\mathrm{U}}) &= \mathrm{softmax}\left(\frac{Q_{\mathrm{U},p} {K_{\mathrm{U},p}} ^{\top}}{\sqrt{d_{h_u}/P}} \right)V_{\mathrm{U},p}, \\
        Q_{\mathrm{U},p} & = H_{\mathrm{U}} W_{\mathrm{U},p}^Q, K_{\mathrm{U},p} = H_{\mathrm{U}} W_{\mathrm{U},p}^K, V_{\mathrm{U},p} = H_{\mathrm{U}} W_{\mathrm{U},p}^V
    \end{split}
\end{equation}
here $W_*^Q, W_*^K, W_*^V$, and $W_*^O$ are trainable parameters. 
The multi-head attention layer takes users' separate nodes representations as input and exploits the high-order connections with the self-attention mechanism. For each domain, the corresponding node can adaptively refine its preference representation by extracting the correlative information from other domains.

\subsection{Model Optimization and Time Complexity}\label{hypergraph_train}
These two hyperedge-base modules: dynamic item transfer module (Hyper-I) and adaptive user aggregation module (Hyper-U), compose a hierarchical hypergraph neural network. It realizes correlative preference transfer and exploits the high-order connection among users' multi-domain behaviors. 

For model optimization, we mix the multi-domain data and randomly select a sample $(u^m, i)$ from domain $\mathcal D_m$ for each training step. Domain $\mathcal D_m$ is taken as the target domain and the others are source domains. We employ a contrastive loss InfoNCE~\cite{van2018representation} to optimize the model, which maximizes the agreements between positive pairs. Formally,
\begin{equation}
    \mathcal L(u, i\mid \mathcal D_m) = - \log \frac{\exp (sim(z_{u^m}, z_i)/\tau)}{\sum_{i_{-}}  \exp(sim(z_{u^m}, z_{i_{-}})/\tau)} 
\end{equation}
where $sim(\cdot)$ stands for similarity measure function and
we use inner product. $(u^m, i_{-})$ is a randomly sampled negative pair that $r^m_{u, i_{-}} = 0$, and $\tau$ is the temperature hyperparameter.

The time complexity of Hyper-U module is $\mathcal{O}(T^2d+Td^2)$, where $T$ is domain number and $d$ is embedding dim. 
For Hyper-I module, the time complexity is $\mathcal{O}(Tn(k^2d+kd^2))$, where $n$ is the number of sampled neighbors and  $k$ is the size of similar item set. The main limitation of  $\mathsf{H^3Trans}$ is computation cost and memory cost (incorporating hyperedges). Compared to the baselines that trains models for multiple domains in parallel, $\mathsf{H^3Trans}$ unifies all domain data and training time increases. In future work, we shall focus on efficient algorithms, i.e., reducing memory cost via hyperedge dropout and reducing time complexity via accelerating self-attention.

\section{Experiments}
In this section, we conduct both offline and online experiments to validate the effectiveness of our method. And the experiments are intended to answer the following research questions:
\begin{itemize}
    \item \textbf{RQ1}: How does our proposed method perform when compared with other state-of-the-art GNN-based methods?
    \item \textbf{RQ2}: How do the different components (e.g., unified multi-domain graph, adaptive user aggregation module, dynamic item transfer module) contribute to the model performance?
    \item \textbf{RQ3}: Does our method help alleviate the behavior sparseness issue and improve recommendation performance for the relatively inactive users (with fewer interaction items)?
    \item \textbf{RQ4}: Does $\mathsf{H^3Trans}$ achieve improvement when deployed to our advertising system?
\end{itemize}

\begin{table}[t]
    \centering
    \vspace{-1em}
    \setlength{\abovecaptionskip}{0.2cm}
    \footnotesize
    \caption{Dataset Statistics}
    \begin{tabular}{cccc|cccc}
        \toprule
        \multicolumn{4}{c|}{Product Dataset} & \multicolumn{4}{c}{Public Amazon Dataset} \\
        Domains & \#user & \#item & \#click & Domains & \#user & \#item & \#click \\
        \midrule
        MDR-A & 84.6M & 6.3M & 3.1B & Books & 1.67M & 0.99M & 26.8M \\
        MDR-B & 34.0M & 1.4M & 0.6B & Music & 0.11M & 0.12M & 1.5M \\
        MDR-C & 24.7M & 0.5M & 0.3B & Movie & 0.23M & 0.08M & 3.1M \\
        MDR-D & 29.1M & 0.6M & 0.2B & - & - & - & - \\
        \bottomrule
    \end{tabular}
    \label{tab:dataset}
\end{table}

\subsection{Experimental Settings}

\subsubsection{\textbf{Datasets}}
We conduct extensive offline experiments on both the public dataset and the product dataset.

\textbf{Public Dataset:} 
Amazon dataset~\cite{ni2019justifying} is a popular dataset to conduct experiments of multi-domain recommendation. The dataset provides dozens of domains and the frequently-used domains are Books, Movies and TV (Movie), and CDS and Vinyl (Music). Following existing research, we take binarize the ratings to 1 and 0 (the ratings higher or equal to 4 as positive and others as negative.) Besides, we filter the users and items with less than 5 interactions. 

\textbf{Product Dataset:} 
The product dataset is collected from four real-world scenarios from an industry advertising platform, named MDR-A, MDR-B, MDR-C, and MDR-D.
These four sub-datasets share the same user set and have overlapped items. Each subset consists of users' interacted items. We additionally filter the datasets to retain users/items with at least 5 interactions. Table~\ref{tab:dataset} lists the statistics of both the product dataset and the public amazon dataset.

\begin{table*}[t]
    \centering
    \setlength{\abovecaptionskip}{0.1cm}
    \caption{Main results on product dataset}
    \begin{tabular}{c|ccc|ccc|ccc|ccc}
        \toprule
        \multirow{2}{*}{Method} & \multicolumn{3}{c|}{MDR-A} & \multicolumn{3}{c|}{MDR-B} & \multicolumn{3}{c|}{MDR-C} & \multicolumn{3}{c}{MDR-D} \\
         & Mrr & HR@20 & HR@50 & Mrr & HR@20 & HR@50 & Mrr & HR@20 & HR@50 & Mrr & HR@20 & HR@50 \\
        \midrule
        Base   & 0.0368 & 2.37\% & 6.46\% & 0.0625 & 4.87\% & 12.60\% & 0.0640 & 4.78\% & 13.20\% & 0.0753 & 5.11\% & 12.65\% \\
        PPGN (Mix)  & 0.0481 & 2.98\% & 8.47\% & 0.0603 & 4.22\% & 11.61\% & 0.1017 & 8.58\% & 18.99\% & 0.1131 & 7.60\% & 17.59\% \\
        MGNN    & 0.0544 & 3.68\% & 8.11\% & 0.0699  & 5.34\% & 14.28\% & 0.1079 & 12.22\% & 21.34\% & 0.1428 & 10.67\% & 21.81\%   \\
        PCRec   & 0.0635 & 4.38\% & 9.71\% & 0.0845 & 7.31\% & 16.63\% & 0.1546 & 14.71\% & 25.99\% & 0.1738 & 15.16\% & 26.59\% \\
        BiTGCF  & 0.0663 & 4.59\% & 10.61\% & 0.0986 & 8.66\% & 18.46\% & 0.1591 & 15.48\% & 26.49\% & 0.1577 & 13.73\% & 23.66\% \\
        BiTGCF+  & 0.0750 & 5.08\% & 12.31\% & 0.1237 & 9.87\% & 20.71\% & 0.1713 & 16.15\% & 28.63\% & 0.1685 & 14.76\% & 25.85\% \\ 
        \midrule
        $\mathsf{H^3Trans}$ & \textbf{0.1171} & \textbf{7.20\%} & \textbf{16.79\%} & \textbf{0.1686} &  \textbf{14.29\%} & \textbf{28.65\%} & \textbf{0.2084} & \textbf{18.78\%} & \textbf{34.89\%} & \textbf{0.2158} & \textbf{18.69\%} & \textbf{32.73\%}  \\
        \bottomrule
    \end{tabular}
    \label{tab:main_results}
    \vspace{-0.6em}
\end{table*}

\begin{table}[t]
    \centering
    \small
    \setlength{\abovecaptionskip}{0.1cm}
    \caption{Main results on public amazon dataset}
    \begin{tabular}{c|cc|cc|cc}
        \toprule
        \multirow{2}{*}{Method} & \multicolumn{2}{c|}{Books} & \multicolumn{2}{c|}{Music} & \multicolumn{2}{c}{Movie}  \\
         & NDCG & HR@20 & NDCG & HR@20 & NDCG & HR@20 \\
        \midrule
        Base & 0.0270 & 4.71\% & 0.0631 & 13.39\% & 0.0433 & 10.45\% \\
        PPGN & 0.0289 & 4.96\% & 0.0660 & 13.93\% & 0.0473 & 11.23\% \\ 
        MGNN & 0.0311 & 5.12\% & 0.0672 & 14.14\% & 0.0465 & 11.03\% \\
        PCRec & 0.0331 & 5.31\% & 0.0742 & 15.67\% & 0.0489 & 11.52\% \\
        BitGCF & 0.0359 & 5.57\% & 0.0694 & 14.65\% & 0.0495 & 11.78\% \\
        BitGCF+ & 0.0381 & 5.78\% & 0.0719 & 15.29\% & 0.0509 & 12.02\% \\
        \midrule
        $\mathsf{H^3Trans}$ & \textbf{0.0399} & \textbf{5.97\%} & \textbf{0.0761} & \textbf{16.01\%} & \textbf{0.0524} & \textbf{12.33\%}\\
        \bottomrule
    \end{tabular}
    \label{tab:public_main_results}
    \vspace{-1em}
\end{table}

\subsubsection{\textbf{Compared methods}}
We compare $\mathsf{H^3Trans}$ with following strong baselines. Except for the base model, all baselines attempt to transfer information from other domains in different ways. 
\begin{itemize}
    \item \textbf{Base}. Base method constructs a user-item bipartite graph and trains models individually for each domain with its user behavior data.  
    \item \textbf{PPGN}. PPGN~\cite{PPGN} fuses the interaction information of multiple domains into a graph and shares the features of users learned from the joint interaction graph. Notes that one user only has one node within the joint graph. 
    \item \textbf{MGNN}. MGNN~\cite{MGNN} integrates users' multi-domain behaviors and constructs the unified multi-domain graph. Nodes belonging to the same user share the same attribute.  
    MGNN learns domain-specific representation for user nodes. 
    \item \textbf{PCRec}. PCRec~\cite{pretrain} adopts a pre-training and fine-tuning diagram to transfer knowledge from the source domain to the target. Here we first pre-train a graph model on the joint graph and then fine-tune it on each domain. 
    \item \textbf{BiTGCF}. BiTGCF~\cite{BiTGCF} is proposed for dual-target recommendation. It connects common users of both domains as bridge and designs a feature transfer layer to realize the two-way transfer of knowledge across two domains. Here we randomly pick two domains to realize the combination layer. 
    \item \textbf{BiTGCF+}. BiTGCF+ is an extended version of BiTGCF. Here we modify the feature transfer layer and extend it to multi-domain recommendation. 
\end{itemize}

\subsubsection{\textbf{Evaluation Protocol}}
We adopt the widely used leave-one-out evaluation method. Specifically, we take the last interaction from each user’s interaction history as the test set, and the remaining are utilized for training. 
For users in the test set, we follow the all-ranking protocol ~\cite{wang2019neural} to evaluate the top-K recommendation performance. For product dataset, we report the average HitRate@K (HR@K) and Mean Reciprocal Rank (MRR) on each domain. For public dataset, we report the HR@K and NDCG@K as these two metrics are more popular of public experiments.

\subsubsection{\textbf{Implementation Details}}
We provide the implementation details of our proposed model and baselines. For fair comparison, each of graph neural network models has two layers, and the hidden embedding dimensions are set as [128, 64]. We sample $k=20$ related items to build \textit{hyperedge-i} in Hyper-I module. For model training, we set batch size $N = 512$ and adopt adam optimizer~\cite{kingma2014adam}, where the learning rate is set to $0.01$.

\subsection{Performance Comparison (RQ1)}
Table~\ref{tab:main_results} and Table~\ref{tab:public_main_results} present the experimental results of $\mathsf{H^3Trans}$ compared with other baselines. From these two tables, we have the following observations.
\begin{itemize}[leftmargin=*]
    \item Base method performs poorly on all domains, which indicates that individually training model for each domain limits the recommendation performance in multi-domain recommendation. 
    \item PPGN mixes the multi-domain data and constructs a joint graph for model training. As a result, it achieves large improvement in most domains. But it still has negative effects on some domains such as MDR-B, because different domains share the same user representation and neglect the user's domain-specific preferences. The user representation is dominated by the data-rich domain.
    \item MGNN takes account of both the common feature and the domain-specific feature for different domains. which brings improvement to the recommendation service. Note that common feature is only acquired by the shared node attributes. The information transfer among domains is limited.
    \item PCRec performs transfer learning by adopting the pre-training and fine-tuning diagram. Pre-training on the joint graph helps learn users' common preferences among domains. Then fine-tuning on domain's individual graph make the user node representation more preferable for each domain. However, fine-tuning is more time- and space-consuming for multi-domain recommendation. 
    \item BiTGCF and BiTGCF+ are two competitive baselines in our experiments. BiTGCF leverages a combination layer to realize the two-way transfer across domains. Here we extend the feature transfer layer of BiTGCF to multiple domains as BiTGCF+. We can see that BitGCF+ achieves larger improvement than BitGCF because it introduces more domains to perform multi-domain recommendation. But the improvement is still limited because we just simply sum user's multi-domain representations and neglect the high-order connections among them.
    \item $\mathsf{H^3Trans}$ achieves the best performance with significant improvement on all metrics of all domains. This indicates that $\mathsf{H^3Trans}$ benefits from learning the high-order connections among multiple domains extracted by Hyper-U module and transferring correlative information via Hyper-I. The high-quality representations learned from the hypergraph enhance the recommendation performance in all domains. 
\end{itemize}

\subsection{Ablation Study (RQ2)}
For further analysis, we compare different variants of $\mathsf{H^3Trans}$ on the product dataset for ablation study, and the results are listed in table~\ref{tab:ablation_study}. Vanilla is a basic graph model trained on the unified multi-domain graph. User nodes learn the common interest only through the shared node attributes. 

\subsubsection{\textbf{Effect of Hyper-U module}:}
HU adds the Hyper-U module but without the attention mechanism based layer, which is equivalent to BiTGCF+. It only utilizes a vanilla combination layer to combine users' separate representations from multiple domains. HU+ integrates our self-attention mechanism based message passing layer into HU. From the table, we can see that aggregating users' scattered preferences and modeling the high-order connections among multiple domains could help refine the user representation for each separate domain. And the self-attention mechanism contributes to further improving the representation quality, because the attention layer adaptively extracts correlative knowledge from source domains.  

\subsubsection{\textbf{Effect of Hyper-I module}:}
PHI and EHI are two models that additionally integrate the Hyper-I module, and equipped with path-based or embed-based method to seek out similar items respectively. Table~\ref{tab:ablation_study} shows that these two methods perform better than HU+, which indicates that the dynamic item transfer module could eliminate the domain discrepancy and adjust the latent item representation more correlative to the target domain without interference information. 
Besides, EHI achieves a marginal improvement than PHI, that shows embed-based method is a little better than path-based method. EHI+ is the best variant of our model, which further employs the graphormer layer to exploit the structure information within the {hyperedge-i}. It consistently shows around 1\% on HR@20 and 2\% on HR@50. 

The itemset size of each domain ranges from tens-of-thousands to millions, while the size of selected correlative itemset is $K$. The value of $K$ is a key hyperparameter: A too small value brings unstable training. A too large value increases computation cost, and different source items usually retrieve similar itemsets that lacks of discriminatory information.

\begin{figure}[t]
\centering
\centerline{\includegraphics[width=\columnwidth]{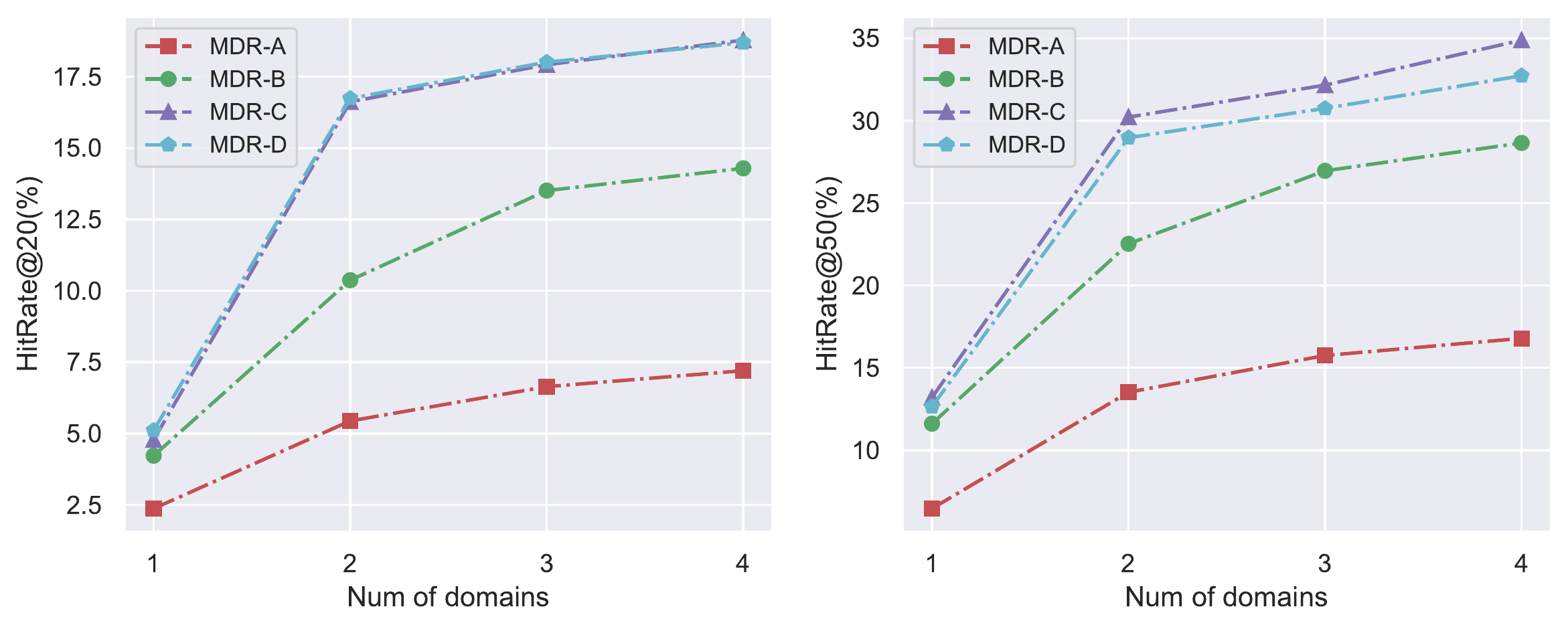}}
\vspace{-1em}
\caption{Performance comparison over different number of domains in MDR}
\label{fig:domains_num}
\vspace{-1em}
\end{figure}

\subsubsection{\textbf{Effect of multiple domains}:}
Multi-domain recommendation jointly optimizes the recommendation performance of all domains. Intuitively, with more domains, we can access more users' behaviors to better characterize users' interest. Here we analyze the effect when introducing different numbers of domains to perform multi-domain recommendation. The results are reported in figure~\ref{fig:domains_num}. We can see that it indeed achieves better performance when introducing more domains, because we can transfer knowledge from more source domains, and $\mathsf{H^3Trans}$ help exploit the high-order connections among them. Additionally, the marginal improvement decreases as more domains are introduced.

\subsection{Alleviating  Behavior Sparseness (RQ3)}
\begin{figure}[t]
\centering
\centerline{\includegraphics[width=\columnwidth]{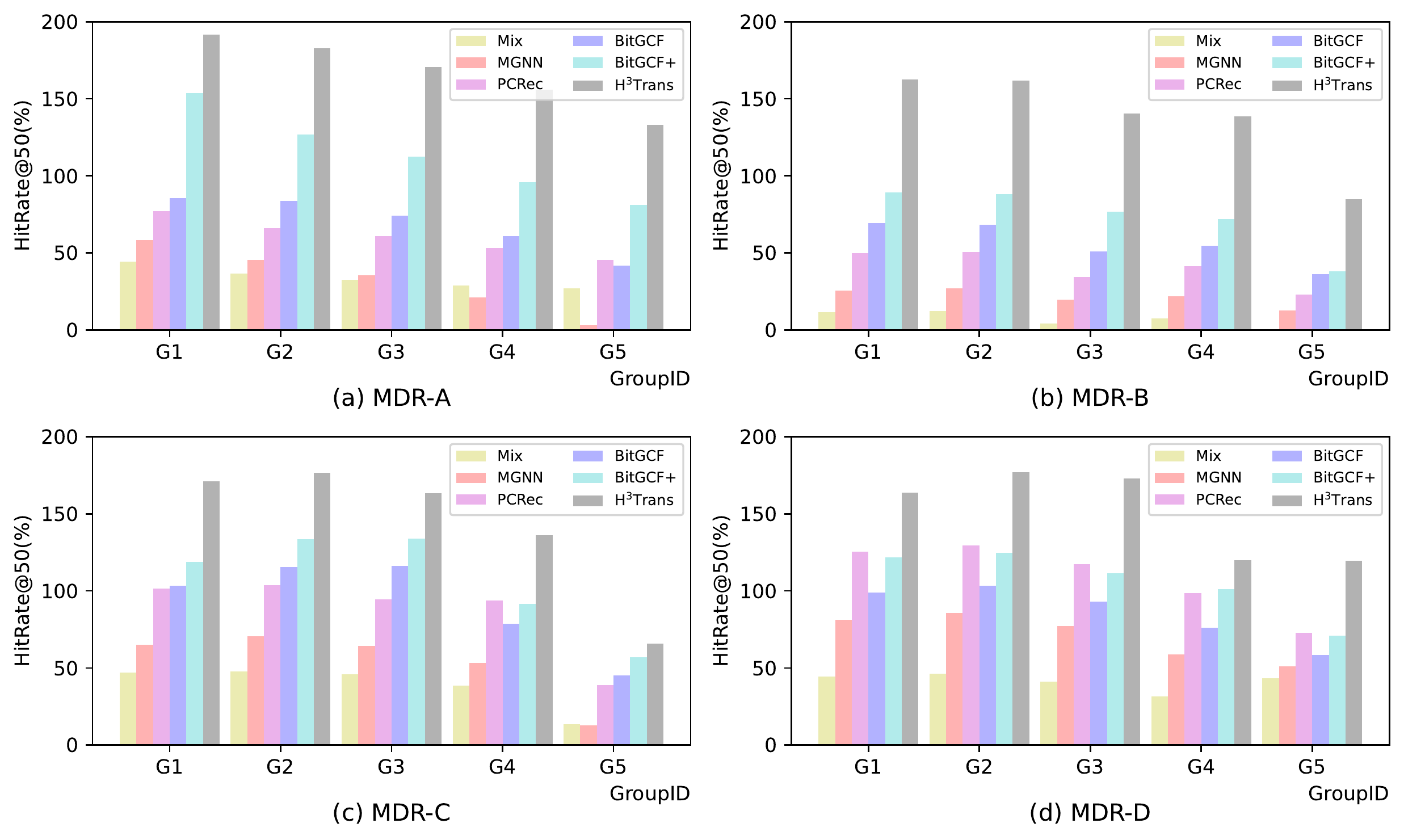}}
\vspace{-1em}
\caption{Performance comparison over different user groups (percentage increase relative to Base model)}
\label{fig:sparsity_analysis}
\vspace{-1.6em}
\end{figure}

As stated before, GNN-based methods suffer from the behavior sparseness issue, and here we conduct a detailed analysis to test the improvement on behavior-sparse users. Specifically, we split the users into four groups G1, G2, G3, G4, and G5 in the order of increasing number of interactions. The larger the GroupID is, the more behaviors the users have collected. Figure~\ref{fig:sparsity_analysis} reports the percentage increase compared with the Base model. We can find that the improvement achieved in the first three groups is more significant than that of the last two. We conclude that $\mathsf{H^3Trans}$ help improve more for relatively inactive users (with fewer user-item interactions), indicating that $\mathsf{H^3Trans}$ alleviates the sparseness of user behaviors by transferring knowledge from other domains.

\begin{table*}[t]
    \centering
    \setlength{\abovecaptionskip}{0.2cm}
    \caption{Ablation study on product dataset. Methods refer to different variants of $\mathsf{H^3Trans}$.}
    \begin{tabular}{c|ccc|ccc|ccc|ccc}
        \toprule
        \multirow{2}{*}{Method} &  \multicolumn{3}{c|}{MDR-A} & \multicolumn{3}{c|}{MDR-B} & \multicolumn{3}{c|}{MDR-C} & \multicolumn{3}{c}{MDR-D} \\
         & Mrr & HR@20 & HR@50 & Mrr & HR@20 & HR@50 & Mrr & HR@20 & HR@50 & Mrr & HR@20 & HR@50 \\
        \midrule
        Vanilla & 0.0544 & 3.68\% & 8.11\% & 0.0699  & 5.34\% & 14.28\% & 0.1079 & 12.22\% & 21.34\% & 0.1428 & 10.67\% & 21.81\%   \\
        HU  & 0.0750 & 5.08\% & 12.31\% & 0.1237 & 9.87\% & 20.71\% & 0.1712 & 16.15\% & 28.63\% & 0.1685 & 14.76\% & 25.85\% \\ 
        HU+ & 0.0894 & 5.56\% & 13.68\% & 0.1383 & 10.53\% & 23.08\% & 0.1848 & 17.01\% & 29.82\% & 0.1846  & 16.38\% & 28.48\% \\
        PHI & 0.1016 & 6.35\% & 15.22\% & 0.1509 & 11.96\% & 24.52\% & 0.1887 & 17.58\% & 30.92\% & 0.1913 & 17.21\% & 29.80\% \\
        EHI & 0.1051 & 6.53\% & 15.68\% & 0.1581 & 12.34\% & 25.54\% & 0.1958 & 17.93\% & 31.64\% & 0.1937 & 17.84\% & 30.62\%  \\
        EHI+ & \textbf{0.1171} & \textbf{7.20\%} & \textbf{16.79\%} & \textbf{0.1686} &  \textbf{14.29\%} & \textbf{28.65\%} & \textbf{0.2084} & \textbf{18.78\%} & \textbf{34.89\%} & \textbf{0.2158} & \textbf{18.69\%} & \textbf{32.73\%}  \\
        \bottomrule
    \end{tabular}
    \vspace{-0.5em}
    \label{tab:ablation_study}
\end{table*}

\subsection{Online Experiment (RQ4)}
We have deployed $\mathsf{H^3Trans}$ online to the retrieval module of our advertising system for an emerging scenario, and conducted online A/B test for one week. For fair comparison, we follow the same configuration with the best retrieval model deployed online~\cite{pretrain}. The online metrics include CTR, conversion rate (CVR), gross merchandise volume (GMV) and return on investment (ROI).

We observe that $\mathsf{H^3Trans}$ achieves \textbf{+2.8\%} lift on CTR, \textbf{+10.9\%} lift on CVR, \textbf{+6.7\%} lift on GMV and \textbf{+7.3\%} lift on ROI, and the daily improvement over baseline is stable. The uplift is mainly from  users having lowest activity level, verifying that $\mathsf{H^3Trans}$ learns high-quality embeddings for inactive users through preference transfer. Therefore $\mathsf{H^3Trans}$ improves the important online metrics and promotes the performance to our system.

\section{Related Work}

\subsection{Multi-domain Recommendation}
Multi-domain recommendation aims to improve recommendations performance of all domains by transferring knowledge from related domains. MCF~\cite{zhang2010multi} and ICAN~\cite{xie2020internal} consider multiple collaborative filtering tasks in different domains simultaneously and exploit the relationships between domains.
~\citet{ma2018your} further introduce cross-media content information.
Some works focus on the users' multiple behaviors. MBGCN ~\cite{jin2020multi} and MGNN~\cite{MGNN} propose a multi-behavior graph convolutional network to capture behaviors' different influences on target behavior. Furthermore, by considering each domain as a task, multi-task approaches can be directly applied in MDR. For general MDR, MMoE~\cite{ma2018modeling} models the tradeoffs between domain-specific objectives and inter-domain relationships with a new multi-gate expert strategy.

\subsection{GNNs for Cross-domain Recommendation}
Inspired by the success of graph neural networks\cite{hamilton2017inductive,kipf2016semi}, researchers have taken efforts to exploit the user-item interaction behavior graph. GNN-based methods~\cite{he2020lightgcn,wang2019neural,ying2018graph} suffer from the sparseness of user behaviors, and some researchers have exploited to alleviate it by transferring information from other domains~\cite{PPGN,pretrain,BiTGCF}. PPGN~\cite{PPGN} fuses the interaction information of two domains into a graph and learns shared features for users. ~\citet{pretrain} propose a pre-training and fine-tuning diagram to transfer information to the target domain.  ~\citet{BiTGCF}  realizes the two-way transfer of knowledge across two domains with a bi-directional feature transfer module. ~\citet{da-gtcdr} propose a graphical and attentional model to combine the embeddings of common users from both domains, thus enhancing the quality of user embeddings and improving the recommendation performance on each domain. However, they fail to model  high-order connections among more domains. 

\subsection{Hypergraph Learning for Recommendation}
Hypergraph, as a more general topological structure to model high-order connections, has been exploited in recommendation~\cite{ji2020dual,yu2021self,xia2021self,wang2020next,zhang2021double,chen2020neural}. \citet{xia2021self} models session-based data as a hypergraph and then propose a hypergraph convolutional network for session-based recommendation. \citet{yu2021self} propose a multi-channel hypergraph convolutional network to enhance social recommendation by leveraging high-order user connections. \citet{zhang2021double} incorporate the complex tuple-wise correlations into a hypergraph and propose a self-supervised hypergraph learning framework for group recommendation. Our work is the first to investigate hypergraph learning in multi-domain recommendation, which can exploit the high-order connections among multiple domain and realize correlative preference transfer.

\section{Conclusion}
In this paper, we propose an correlative preference transfer framework with hierarchical hypergraph network ($\mathsf{H^3Trans}$) to improve multi-domain recommendations. $\mathsf{H^3Trans}$ constructs a unified multi-domain graph and integrates two hyperedge-based module: adaptive user aggregation and dynamic item transfer.  
$\mathsf{H^3Trans}$ not only exploits high-order connections among users’ scattered preferences in multiple domain, but also transfers correlative user preference to alleviate the behavior sparseness of each single domain. Extensive experiments demonstrate the superiority of our method.

\bibliographystyle{ACM-Reference-Format}
\bibliography{reference.bib}


\begin{thebibliography}{32}


\ifx \showCODEN    \undefined \def \showCODEN     #1{\unskip}     \fi
\ifx \showDOI      \undefined \def \showDOI       #1{#1}\fi
\ifx \showISBNx    \undefined \def \showISBNx     #1{\unskip}     \fi
\ifx \showISBNxiii \undefined \def \showISBNxiii  #1{\unskip}     \fi
\ifx \showISSN     \undefined \def \showISSN      #1{\unskip}     \fi
\ifx \showLCCN     \undefined \def \showLCCN      #1{\unskip}     \fi
\ifx \shownote     \undefined \def \shownote      #1{#1}          \fi
\ifx \showarticletitle \undefined \def \showarticletitle #1{#1}   \fi
\ifx \showURL      \undefined \def \showURL       {\relax}        \fi
\providecommand\bibfield[2]{#2}
\providecommand\bibinfo[2]{#2}
\providecommand\natexlab[1]{#1}
\providecommand\showeprint[2][]{arXiv:#2}

\bibitem[\protect\citeauthoryear{Chen, Pan, Jiang, Huo, and Long}{Chen
  et~al\mbox{.}}{2019}]%
        {dagcn}
\bibfield{author}{\bibinfo{person}{Fengwen Chen}, \bibinfo{person}{Shirui Pan},
  \bibinfo{person}{Jing Jiang}, \bibinfo{person}{Huan Huo}, {and}
  \bibinfo{person}{Guodong Long}.} \bibinfo{year}{2019}\natexlab{}.
\newblock \showarticletitle{DAGCN: dual attention graph convolutional
  networks}. In \bibinfo{booktitle}{\emph{2019 International Joint Conference
  on Neural Networks (IJCNN)}}. IEEE, \bibinfo{pages}{1--8}.
\newblock


\bibitem[\protect\citeauthoryear{Chen, Xiong, Zhang, Xia, Yin, and Huang}{Chen
  et~al\mbox{.}}{2020b}]%
        {chen2020neural}
\bibfield{author}{\bibinfo{person}{Xu Chen}, \bibinfo{person}{Kun Xiong},
  \bibinfo{person}{Yongfeng Zhang}, \bibinfo{person}{Long Xia},
  \bibinfo{person}{Dawei Yin}, {and} \bibinfo{person}{Jimmy~Xiangji Huang}.}
  \bibinfo{year}{2020}\natexlab{b}.
\newblock \showarticletitle{Neural feature-aware recommendation with signed
  hypergraph convolutional network}.
\newblock \bibinfo{journal}{\emph{ACM Transactions on Information Systems
  (TOIS)}} (\bibinfo{year}{2020}).
\newblock


\bibitem[\protect\citeauthoryear{Chen, Wang, Ni, Zeng, and Lin}{Chen
  et~al\mbox{.}}{2020a}]%
        {chen2020scenario}
\bibfield{author}{\bibinfo{person}{Yuting Chen}, \bibinfo{person}{Yanshi Wang},
  \bibinfo{person}{Yabo Ni}, \bibinfo{person}{An-Xiang Zeng}, {and}
  \bibinfo{person}{Lanfen Lin}.} \bibinfo{year}{2020}\natexlab{a}.
\newblock \showarticletitle{Scenario-aware and Mutual-based approach for
  Multi-scenario Recommendation in E-Commerce}. In
  \bibinfo{booktitle}{\emph{2020 International Conference on Data Mining
  Workshops (ICDMW)}}. IEEE, \bibinfo{pages}{127--135}.
\newblock


\bibitem[\protect\citeauthoryear{Chien, Pan, Peng, and Milenkovic}{Chien
  et~al\mbox{.}}{2022}]%
        {Chien2022YouAA}
\bibfield{author}{\bibinfo{person}{Eli Chien}, \bibinfo{person}{Chao Pan},
  \bibinfo{person}{Jianhao Peng}, {and} \bibinfo{person}{Olgica Milenkovic}.}
  \bibinfo{year}{2022}\natexlab{}.
\newblock \showarticletitle{You are AllSet: A Multiset Function Framework for
  Hypergraph Neural Networks}.
\newblock \bibinfo{journal}{\emph{ArXiv}} (\bibinfo{year}{2022}).
\newblock


\bibitem[\protect\citeauthoryear{Gu, Bao, Ou, Li, Cui, Ma, Huang, Liu, and
  Zeng}{Gu et~al\mbox{.}}{2021}]%
        {gu2021self}
\bibfield{author}{\bibinfo{person}{Yulong Gu}, \bibinfo{person}{Wentian Bao},
  \bibinfo{person}{Dan Ou}, \bibinfo{person}{Xiang Li},
  \bibinfo{person}{Baoliang Cui}, \bibinfo{person}{Biyu Ma},
  \bibinfo{person}{Haikuan Huang}, \bibinfo{person}{Qingwen Liu}, {and}
  \bibinfo{person}{Xiaoyi Zeng}.} \bibinfo{year}{2021}\natexlab{}.
\newblock \showarticletitle{Self-Supervised Learning on Users' Spontaneous
  Behaviors for Multi-Scenario Ranking in E-commerce}. In
  \bibinfo{booktitle}{\emph{Proceedings of the 30th ACM International
  Conference on Information \& Knowledge Management}}.
\newblock


\bibitem[\protect\citeauthoryear{Hamilton, Ying, and Leskovec}{Hamilton
  et~al\mbox{.}}{2017}]%
        {hamilton2017inductive}
\bibfield{author}{\bibinfo{person}{Will Hamilton}, \bibinfo{person}{Zhitao
  Ying}, {and} \bibinfo{person}{Jure Leskovec}.}
  \bibinfo{year}{2017}\natexlab{}.
\newblock \showarticletitle{Inductive representation learning on large graphs}.
\newblock \bibinfo{journal}{\emph{Advances in neural information processing
  systems}}  \bibinfo{volume}{30} (\bibinfo{year}{2017}).
\newblock


\bibitem[\protect\citeauthoryear{He, Deng, Wang, Li, Zhang, and Wang}{He
  et~al\mbox{.}}{2020}]%
        {he2020lightgcn}
\bibfield{author}{\bibinfo{person}{Xiangnan He}, \bibinfo{person}{Kuan Deng},
  \bibinfo{person}{Xiang Wang}, \bibinfo{person}{Yan Li},
  \bibinfo{person}{Yongdong Zhang}, {and} \bibinfo{person}{Meng Wang}.}
  \bibinfo{year}{2020}\natexlab{}.
\newblock \showarticletitle{Lightgcn: Simplifying and powering graph
  convolution network for recommendation}. In
  \bibinfo{booktitle}{\emph{Proceedings of the 43rd International ACM SIGIR
  conference on research and development in Information Retrieval}}.
\newblock


\bibitem[\protect\citeauthoryear{Huang and Yang}{Huang and Yang}{2021}]%
        {unignn}
\bibfield{author}{\bibinfo{person}{Jing Huang} {and} \bibinfo{person}{Jie
  Yang}.} \bibinfo{year}{2021}\natexlab{}.
\newblock \showarticletitle{UniGNN: a Unified Framework for Graph and
  Hypergraph Neural Networks}. In \bibinfo{booktitle}{\emph{Proceedings of the
  Thirtieth International Joint Conference on Artificial Intelligence
  (IJCAI-21)}}.
\newblock


\bibitem[\protect\citeauthoryear{Ji, Feng, Ji, Zhao, Tang, and Gao}{Ji
  et~al\mbox{.}}{2020}]%
        {ji2020dual}
\bibfield{author}{\bibinfo{person}{Shuyi Ji}, \bibinfo{person}{Yifan Feng},
  \bibinfo{person}{Rongrong Ji}, \bibinfo{person}{Xibin Zhao},
  \bibinfo{person}{Wanwan Tang}, {and} \bibinfo{person}{Yue Gao}.}
  \bibinfo{year}{2020}\natexlab{}.
\newblock \showarticletitle{Dual channel hypergraph collaborative filtering}.
  In \bibinfo{booktitle}{\emph{Proceedings of the 26th ACM SIGKDD International
  Conference on Knowledge Discovery \& Data Mining}}.
\newblock


\bibitem[\protect\citeauthoryear{Jin, Gao, He, Jin, and Li}{Jin
  et~al\mbox{.}}{2020}]%
        {jin2020multi}
\bibfield{author}{\bibinfo{person}{Bowen Jin}, \bibinfo{person}{Chen Gao},
  \bibinfo{person}{Xiangnan He}, \bibinfo{person}{Depeng Jin}, {and}
  \bibinfo{person}{Yong Li}.} \bibinfo{year}{2020}\natexlab{}.
\newblock \showarticletitle{Multi-behavior recommendation with graph
  convolutional networks}. In \bibinfo{booktitle}{\emph{Proceedings of the 43rd
  International ACM SIGIR Conference on Research and Development in Information
  Retrieval}}. \bibinfo{pages}{659--668}.
\newblock


\bibitem[\protect\citeauthoryear{Kingma and Ba}{Kingma and Ba}{2014}]%
        {kingma2014adam}
\bibfield{author}{\bibinfo{person}{Diederik~P Kingma} {and}
  \bibinfo{person}{Jimmy Ba}.} \bibinfo{year}{2014}\natexlab{}.
\newblock \showarticletitle{Adam: A method for stochastic optimization}.
\newblock \bibinfo{journal}{\emph{arXiv preprint arXiv:1412.6980}}
  (\bibinfo{year}{2014}).
\newblock


\bibitem[\protect\citeauthoryear{Kipf and Welling}{Kipf and Welling}{2016}]%
        {kipf2016semi}
\bibfield{author}{\bibinfo{person}{Thomas~N Kipf} {and} \bibinfo{person}{Max
  Welling}.} \bibinfo{year}{2016}\natexlab{}.
\newblock \showarticletitle{Semi-supervised classification with graph
  convolutional networks}.
\newblock \bibinfo{journal}{\emph{arXiv preprint arXiv:1609.02907}}
  (\bibinfo{year}{2016}).
\newblock


\bibitem[\protect\citeauthoryear{Liu, Li, Li, and Pan}{Liu
  et~al\mbox{.}}{2020}]%
        {BiTGCF}
\bibfield{author}{\bibinfo{person}{Meng Liu}, \bibinfo{person}{Jianjun Li},
  \bibinfo{person}{Guohui Li}, {and} \bibinfo{person}{Peng Pan}.}
  \bibinfo{year}{2020}\natexlab{}.
\newblock \showarticletitle{Cross Domain Recommendation via Bi-Directional
  Transfer Graph Collaborative Filtering Networks}. In
  \bibinfo{booktitle}{\emph{Proceedings of the 29th ACM International
  Conference on Information \& Knowledge Management}}.
\newblock


\bibitem[\protect\citeauthoryear{Ma, Zhao, Yi, Chen, Hong, and Chi}{Ma
  et~al\mbox{.}}{2018b}]%
        {ma2018modeling}
\bibfield{author}{\bibinfo{person}{Jiaqi Ma}, \bibinfo{person}{Zhe Zhao},
  \bibinfo{person}{Xinyang Yi}, \bibinfo{person}{Jilin Chen},
  \bibinfo{person}{Lichan Hong}, {and} \bibinfo{person}{Ed~H Chi}.}
  \bibinfo{year}{2018}\natexlab{b}.
\newblock \showarticletitle{Modeling task relationships in multi-task learning
  with multi-gate mixture-of-experts}. In \bibinfo{booktitle}{\emph{Proceedings
  of the 24th ACM SIGKDD international conference on knowledge discovery \&
  data mining}}. \bibinfo{pages}{1930--1939}.
\newblock


\bibitem[\protect\citeauthoryear{Ma, Zhang, Wang, Luo, Liu, and Ma}{Ma
  et~al\mbox{.}}{2018a}]%
        {ma2018your}
\bibfield{author}{\bibinfo{person}{Weizhi Ma}, \bibinfo{person}{Min Zhang},
  \bibinfo{person}{Chenyang Wang}, \bibinfo{person}{Cheng Luo},
  \bibinfo{person}{Yiqun Liu}, {and} \bibinfo{person}{Shaoping Ma}.}
  \bibinfo{year}{2018}\natexlab{a}.
\newblock \showarticletitle{Your Tweets Reveal What You Like: Introducing
  Cross-media Content Information into Multi-domain Recommendation.}. In
  \bibinfo{booktitle}{\emph{IJCAI}}. \bibinfo{pages}{3484--3490}.
\newblock


\bibitem[\protect\citeauthoryear{Ni, Li, and McAuley}{Ni et~al\mbox{.}}{2019}]%
        {ni2019justifying}
\bibfield{author}{\bibinfo{person}{Jianmo Ni}, \bibinfo{person}{Jiacheng Li},
  {and} \bibinfo{person}{Julian McAuley}.} \bibinfo{year}{2019}\natexlab{}.
\newblock \showarticletitle{Justifying recommendations using distantly-labeled
  reviews and fine-grained aspects}. In \bibinfo{booktitle}{\emph{Proceedings
  of the 2019 conference on empirical methods in natural language processing
  and the 9th international joint conference on natural language processing
  (EMNLP-IJCNLP)}}. \bibinfo{pages}{188--197}.
\newblock


\bibitem[\protect\citeauthoryear{Van~den Oord, Li, and Vinyals}{Van~den Oord
  et~al\mbox{.}}{2018}]%
        {van2018representation}
\bibfield{author}{\bibinfo{person}{Aaron Van~den Oord}, \bibinfo{person}{Yazhe
  Li}, {and} \bibinfo{person}{Oriol Vinyals}.} \bibinfo{year}{2018}\natexlab{}.
\newblock \showarticletitle{Representation learning with contrastive predictive
  coding}.
\newblock \bibinfo{journal}{\emph{arXiv e-prints}} (\bibinfo{year}{2018}).
\newblock


\bibitem[\protect\citeauthoryear{Vaswani, Shazeer, Parmar, Uszkoreit, Jones,
  Gomez, Kaiser, and Polosukhin}{Vaswani et~al\mbox{.}}{2017}]%
        {vaswani2017attention}
\bibfield{author}{\bibinfo{person}{Ashish Vaswani}, \bibinfo{person}{Noam
  Shazeer}, \bibinfo{person}{Niki Parmar}, \bibinfo{person}{Jakob Uszkoreit},
  \bibinfo{person}{Llion Jones}, \bibinfo{person}{Aidan~N Gomez},
  \bibinfo{person}{{\L}ukasz Kaiser}, {and} \bibinfo{person}{Illia
  Polosukhin}.} \bibinfo{year}{2017}\natexlab{}.
\newblock \showarticletitle{Attention is all you need}.
\newblock \bibinfo{journal}{\emph{Advances in neural information processing
  systems}}  \bibinfo{volume}{30} (\bibinfo{year}{2017}).
\newblock


\bibitem[\protect\citeauthoryear{Wang, Zhang, Zhang, Lyu, and Tang}{Wang
  et~al\mbox{.}}{2020b}]%
        {DAN}
\bibfield{author}{\bibinfo{person}{Bei Wang}, \bibinfo{person}{Chenrui Zhang},
  \bibinfo{person}{Hao Zhang}, \bibinfo{person}{Xiaoqing Lyu}, {and}
  \bibinfo{person}{Zhi Tang}.} \bibinfo{year}{2020}\natexlab{b}.
\newblock \showarticletitle{Dual Autoencoder Network with Swap Reconstruction
  for Cold-Start Recommendation}. In \bibinfo{booktitle}{\emph{Proceedings of
  the 29th ACM International Conference on Information \& Knowledge
  Management}}. \bibinfo{pages}{2249--2252}.
\newblock


\bibitem[\protect\citeauthoryear{Wang, Liang, Liu, Zhang, and Yu}{Wang
  et~al\mbox{.}}{2021}]%
        {pretrain}
\bibfield{author}{\bibinfo{person}{Chen Wang}, \bibinfo{person}{Yueqing Liang},
  \bibinfo{person}{Zhiwei Liu}, \bibinfo{person}{Tao Zhang}, {and}
  \bibinfo{person}{Philip~S. Yu}.} \bibinfo{year}{2021}\natexlab{}.
\newblock \showarticletitle{Pre-training Graph Neural Network for Cross Domain
  Recommendation}.
\newblock \bibinfo{journal}{\emph{CoRR}}  \bibinfo{volume}{abs/2111.08268}
  (\bibinfo{year}{2021}).
\newblock
\showeprint[arXiv]{2111.08268}


\bibitem[\protect\citeauthoryear{Wang, Ding, Hong, Liu, and Caverlee}{Wang
  et~al\mbox{.}}{2020a}]%
        {wang2020next}
\bibfield{author}{\bibinfo{person}{Jianling Wang}, \bibinfo{person}{Kaize
  Ding}, \bibinfo{person}{Liangjie Hong}, \bibinfo{person}{Huan Liu}, {and}
  \bibinfo{person}{James Caverlee}.} \bibinfo{year}{2020}\natexlab{a}.
\newblock \showarticletitle{Next-item recommendation with sequential
  hypergraphs}. In \bibinfo{booktitle}{\emph{Proceedings of the 43rd
  international ACM SIGIR conference on research and development in information
  retrieval}}.
\newblock


\bibitem[\protect\citeauthoryear{Wang, He, Wang, Feng, and Chua}{Wang
  et~al\mbox{.}}{2019}]%
        {wang2019neural}
\bibfield{author}{\bibinfo{person}{Xiang Wang}, \bibinfo{person}{Xiangnan He},
  \bibinfo{person}{Meng Wang}, \bibinfo{person}{Fuli Feng}, {and}
  \bibinfo{person}{Tat-Seng Chua}.} \bibinfo{year}{2019}\natexlab{}.
\newblock \showarticletitle{Neural graph collaborative filtering}. In
  \bibinfo{booktitle}{\emph{Proceedings of the 42nd international ACM SIGIR
  conference on Research and development in Information Retrieval}}.
  \bibinfo{pages}{165--174}.
\newblock


\bibitem[\protect\citeauthoryear{Xia, Yin, Yu, Wang, Cui, and Zhang}{Xia
  et~al\mbox{.}}{2021}]%
        {xia2021self}
\bibfield{author}{\bibinfo{person}{Xin Xia}, \bibinfo{person}{Hongzhi Yin},
  \bibinfo{person}{Junliang Yu}, \bibinfo{person}{Qinyong Wang},
  \bibinfo{person}{Lizhen Cui}, {and} \bibinfo{person}{Xiangliang Zhang}.}
  \bibinfo{year}{2021}\natexlab{}.
\newblock \showarticletitle{Self-supervised hypergraph convolutional networks
  for session-based recommendation}. In \bibinfo{booktitle}{\emph{Proceedings
  of the AAAI Conference on Artificial Intelligence}}.
\newblock


\bibitem[\protect\citeauthoryear{Xie, Qiu, Rao, Liu, Zhang, and Lin}{Xie
  et~al\mbox{.}}{2020}]%
        {xie2020internal}
\bibfield{author}{\bibinfo{person}{Ruobing Xie}, \bibinfo{person}{Zhijie Qiu},
  \bibinfo{person}{Jun Rao}, \bibinfo{person}{Yi Liu}, \bibinfo{person}{Bo
  Zhang}, {and} \bibinfo{person}{Leyu Lin}.} \bibinfo{year}{2020}\natexlab{}.
\newblock \showarticletitle{Internal and Contextual Attention Network for
  Cold-start Multi-channel Matching in Recommendation.}. In
  \bibinfo{booktitle}{\emph{IJCAI}}. \bibinfo{pages}{2732--2738}.
\newblock


\bibitem[\protect\citeauthoryear{Ying, Cai, Luo, Zheng, Ke, He, Shen, and
  Liu}{Ying et~al\mbox{.}}{2021}]%
        {graphormer}
\bibfield{author}{\bibinfo{person}{Chengxuan Ying}, \bibinfo{person}{Tianle
  Cai}, \bibinfo{person}{Shengjie Luo}, \bibinfo{person}{Shuxin Zheng},
  \bibinfo{person}{Guolin Ke}, \bibinfo{person}{Di He},
  \bibinfo{person}{Yanming Shen}, {and} \bibinfo{person}{Tie-Yan Liu}.}
  \bibinfo{year}{2021}\natexlab{}.
\newblock \showarticletitle{Do transformers really perform badly for graph
  representation?}
\newblock \bibinfo{journal}{\emph{Advances in Neural Information Processing
  Systems}}  \bibinfo{volume}{34} (\bibinfo{year}{2021}),
  \bibinfo{pages}{28877--28888}.
\newblock


\bibitem[\protect\citeauthoryear{Ying, He, Chen, Eksombatchai, Hamilton, and
  Leskovec}{Ying et~al\mbox{.}}{2018}]%
        {ying2018graph}
\bibfield{author}{\bibinfo{person}{Rex Ying}, \bibinfo{person}{Ruining He},
  \bibinfo{person}{Kaifeng Chen}, \bibinfo{person}{Pong Eksombatchai},
  \bibinfo{person}{William~L Hamilton}, {and} \bibinfo{person}{Jure Leskovec}.}
  \bibinfo{year}{2018}\natexlab{}.
\newblock \showarticletitle{Graph convolutional neural networks for web-scale
  recommender systems}. In \bibinfo{booktitle}{\emph{Proceedings of the 24th
  ACM SIGKDD international conference on knowledge discovery \& data mining}}.
  \bibinfo{pages}{974--983}.
\newblock


\bibitem[\protect\citeauthoryear{Yu, Yin, Li, Wang, Hung, and Zhang}{Yu
  et~al\mbox{.}}{2021}]%
        {yu2021self}
\bibfield{author}{\bibinfo{person}{Junliang Yu}, \bibinfo{person}{Hongzhi Yin},
  \bibinfo{person}{Jundong Li}, \bibinfo{person}{Qinyong Wang},
  \bibinfo{person}{Nguyen Quoc~Viet Hung}, {and} \bibinfo{person}{Xiangliang
  Zhang}.} \bibinfo{year}{2021}\natexlab{}.
\newblock \showarticletitle{Self-supervised multi-channel hypergraph
  convolutional network for social recommendation}. In
  \bibinfo{booktitle}{\emph{Proceedings of the Web Conference 2021}}.
\newblock


\bibitem[\protect\citeauthoryear{Zhang, Gao, Yu, Guo, Li, and Yin}{Zhang
  et~al\mbox{.}}{2021}]%
        {zhang2021double}
\bibfield{author}{\bibinfo{person}{Junwei Zhang}, \bibinfo{person}{Min Gao},
  \bibinfo{person}{Junliang Yu}, \bibinfo{person}{Lei Guo},
  \bibinfo{person}{Jundong Li}, {and} \bibinfo{person}{Hongzhi Yin}.}
  \bibinfo{year}{2021}\natexlab{}.
\newblock \showarticletitle{Double-Scale Self-Supervised Hypergraph Learning
  for Group Recommendation}. In \bibinfo{booktitle}{\emph{Proceedings of the
  30th ACM International Conference on Information \& Knowledge Management}}.
\newblock


\bibitem[\protect\citeauthoryear{Zhang, Mao, Cao, and Xu}{Zhang
  et~al\mbox{.}}{2020}]%
        {MGNN}
\bibfield{author}{\bibinfo{person}{Weifeng Zhang}, \bibinfo{person}{Jingwen
  Mao}, \bibinfo{person}{Yi Cao}, {and} \bibinfo{person}{Congfu Xu}.}
  \bibinfo{year}{2020}\natexlab{}.
\newblock \showarticletitle{Multiplex Graph Neural Networks for Multi-Behavior
  Recommendation}. In \bibinfo{booktitle}{\emph{Proceedings of the 29th ACM
  International Conference on Information \& Knowledge Management}}.
\newblock


\bibitem[\protect\citeauthoryear{Zhang, Cao, and Yeung}{Zhang
  et~al\mbox{.}}{2010}]%
        {zhang2010multi}
\bibfield{author}{\bibinfo{person}{Yu Zhang}, \bibinfo{person}{Bin Cao}, {and}
  \bibinfo{person}{Dit-Yan Yeung}.} \bibinfo{year}{2010}\natexlab{}.
\newblock \showarticletitle{Multi-domain collaborative filtering}. In
  \bibinfo{booktitle}{\emph{Proceedings of the Twenty-Sixth Conference on
  Uncertainty in Artificial Intelligence}}. \bibinfo{pages}{725--732}.
\newblock


\bibitem[\protect\citeauthoryear{Zhao, Li, and Fu}{Zhao et~al\mbox{.}}{2019}]%
        {PPGN}
\bibfield{author}{\bibinfo{person}{Cheng Zhao}, \bibinfo{person}{Chenliang Li},
  {and} \bibinfo{person}{Cong Fu}.} \bibinfo{year}{2019}\natexlab{}.
\newblock \showarticletitle{Cross-Domain Recommendation via Preference
  Propagation GraphNet}. In \bibinfo{booktitle}{\emph{Proceedings of the 28th
  ACM International Conference on Information and Knowledge Management}}.
\newblock


\bibitem[\protect\citeauthoryear{Zhu, Wang, Chen, Liu, and Zheng}{Zhu
  et~al\mbox{.}}{2020}]%
        {da-gtcdr}
\bibfield{author}{\bibinfo{person}{Feng Zhu}, \bibinfo{person}{Yan Wang},
  \bibinfo{person}{Chaochao Chen}, \bibinfo{person}{Guanfeng Liu}, {and}
  \bibinfo{person}{Xiaolin Zheng}.} \bibinfo{year}{2020}\natexlab{}.
\newblock \showarticletitle{A Graphical and Attentional Framework for
  Dual-Target Cross-Domain Recommendation.}. In
  \bibinfo{booktitle}{\emph{IJCAI}}. \bibinfo{pages}{3001--3008}.
\newblock


\end{thebibliography}


\end{document}